\documentclass[pre,twocolumn,showpacs]{revtex4-1}
\usepackage{graphicx}

\begin{document}

\title{Jammed systems of oriented needles always percolate on square lattices}
\author{Grzegorz Kondrat}\email{grzegorz.kondrat@uwr.edu.pl}

\author{Zbigniew Koza}
\author{Piotr Brzeski}
\affiliation{Faculty of Physics and Astronomy, University of Wroc{\l}aw, 50-204
Wroc{\l}aw,
Poland}

\date{\today}

\begin{abstract}
Random sequential adsorption (RSA) is a standard method of modeling adsorption
of large molecules at the liquid-solid interface. Several studies have
recently conjectured that in the RSA of rectangular needles, or $k$-mers, on a square
lattice the percolation is impossible if the needles are sufficiently long ($k$~of order of several
thousand). We refute these claims and present a strict proof that in any jammed
configuration of nonoverlapping, fixed-length, horizontal
or vertical needles on a square lattice, all clusters are percolating clusters.
\end{abstract}


\maketitle

%%%%%%%%%%%%%%%%%%%%%%%%%%%%%%%%%%%%%%%%%%%%%%%%%%%%%%%
%%%%%%%%%%%%%%%%%%%%%%%%%%%%%%%%%%%%%%%%%%%%%%%%%%%%%%%
%%%%%%%%%%  Section 1. Introduction   %%%%%%%%%%%%%%%%%
%%%%%%%%%%%%%%%%%%%%%%%%%%%%%%%%%%%%%%%%%%%%%%%%%%%%%%%
%%%%%%%%%%%%%%%%%%%%%%%%%%%%%%%%%%%%%%%%%%%%%%%%%%%%%%%

\section{Introduction\label{sec:intro}}

Adsorption of large molecules like polymers, biomolecules or nanotubes at the
liquid-solid interface is an important part of various natural and technological
processes, including those found in industrial bioreactors \cite{Senger2000}, 
water purification 
\cite{Dabrowski2001}, or production of conducting nanocomposites 
\cite{Mutiso2015}. 
In many cases the adsorption phenomenon is essentially an irreversible and
localized process in which the adsorbed molecules eventually form a monolayer on the
target surface \cite{Feder1980a}. It is therefore quite common to investigate such
phenomena using the random sequential adsorption (RSA) model.
The main idea behind it is simple~\cite{Feder1980a,Evans1993,Talbot2000}:
starting with an
empty substrate, one tries to put on it a sequence of some geometric objects, e.g.\@
disks or rectangles, each at a random position.
An attempt is successful if the new object does not overlap with the ones already
deposited on the surface, otherwise a new attempt is made at
a different, randomly chosen location. Once the object is attached to the
surface, it stays there motionless forever, reducing a chance for the subsequent
molecules to be adsorbed in its neighborhood. The dynamics gradually slows
down and finally no unoccupied room remains on the substrate that could accommodate
the next object---the process stops, the system has reached the so called
jamming limit (a discussion of various aspects of jammed configurations can be found in the review \cite{Torquato2010}).

Numerous extensions of the basic RSA model have been studied so far, including
imperfect substrates \cite{Tarasevich2015,Centres2015},
various object shapes (e.g., disks \cite{Feder1980a}, spheres \cite{Meakin1992},
spherocylinders \cite{Schilling2015},
infinitely thin needles \cite{Provatas2000},
squares \cite{Brosilow1991,Nakamura1986}, ellipses \cite{Viot1992}, and rectangles
\cite{Viot1992,Porto2000,Vandewalle2000,Kondrat2001,Tarasevich2012,
Tarasevich2015,Centres2015}), polydispersity
\cite{Lee2000,Nigro2013,Chatterjee2015},
shape flexibility \cite{Adamczyk2008,Kondrat2002},
post-adsorption dynamics (e.g., desorption \cite{Budinski2001} and
diffusion \cite{Lebovka2017}), and partial \cite{Balberg1987}
or full object overlapping \cite{Torquato2012}.
In each RSA process, basic or extended one, as the molecules are
being deposited onto the surface, they may touch each other
and form larger clusters of connected (or ``neighboring'') objects, and if such
a cluster spans the opposite sides of the system, we have to do with a
percolating cluster.

Consequently, there are two basic quantities characterizing RSA processes.
The first one is the jamming threshold $0 < c_\mathrm{j} \le 1$
defined as the ratio of the surface area covered by the
adsorbed objects ($A_\mathrm{ad}$)
to the total surface area ($A$) in the jammed state.
The second one is the percolation threshold $c_\mathrm{p}$, a quantity similar
to the jamming threshold in that it is also defined as
$A_\mathrm{ad}/A$, except that $A_\mathrm{ad}$ must be now determined
at the moment when the adsorbed molecules start to form a percolating cluster
\cite{Stauffer1994}. While $c_\mathrm{j}$ characterizes any
RSA process, $c_\mathrm{p}$ is well defined only for some of them.
For example, in models where nonoverlapping objects, e.g.\@ circles, are
randomly deposited on a continuous substrate, e.g.\@ a larger
square, no object can actually touch another one and a percolating cluster
cannot be formed.
However, the information whether (or under which conditions) an RSA process leads to
percolation or not is a fundamental characteristic of this process.

In 2000 Vandewalle et al.\@ \cite{Vandewalle2000} advanced a hypothesis that
in the RSA of needle-like rectangles (also known as $k$-mers) on a square lattice
the ratio
$c_\mathrm{p}/c_\mathrm{j}$ is constant for all needle sizes. If correct, this
hypothesis would indicate existence of a deep relation between jamming and
percolation. However, more elaborate studies refuted this claim. Instead,
as will be discussed in detail in Section~\ref{sec:hypothesis}, several
researchers came to the conclusion that for sufficiently long needles the system
does not percolate. However, this striking conjecture is based on extrapolation
of numerical results obtained for relatively short needles and no physical
mechanism responsible for such a behavior is known.

Thus, analysis of the above-mentioned reports
raises the question of percolation breakdown. Does percolation really break down
for very long needles or is it an artifact brought about by using an incorrect
fitting function to the numerical data? There seems to be two ways towards the
solution.
The first one is to carry out a direct numerical examination of
percolation for extremely long needles. However, this would require using
so huge amounts of computer resources (memory and computational time) that such
simulations have not been endeavored yet  \cite{Tarasevich2015}.
The other option is to prove or disprove the conjecture mathematically.
The second option is more attractive, especially since exact arguments in
percolation theory are relatively
rare. 
Here we present such a strict proof that any cluster in a jammed configuration of
fixed-length needles on a square lattice is a percolating cluster. Consequently,
the RSA of fixed-length needles always percolates on a square lattice.

%%%%%%%%%%%%%%%%%%%%%%%%%%%%%%%%%%%%%%%%%%%%%%%%%%%%%%%%%%%%%%%%%%%%%%%%%%%%%%
%%%%%%%%%%%%%%%%%%%%%%%%%%%%%%%%%%%%%%%%%%%%%%%%%%%%%%%%%%%%%%%%%%%%%%%%%%%%%%
%%%%%%%%%%%%%%%%    SECTION 2: Contributions    %%%%%%%%%%%%%%%%%%%%%%%%%%%%%%
%%%%%%%%%%%%%%%%%%%%%%%%%%%%%%%%%%%%%%%%%%%%%%%%%%%%%%%%%%%%%%%%%%%%%%%%%%%%%%
%%%%%%%%%%%%%%%%%%%%%%%%%%%%%%%%%%%%%%%%%%%%%%%%%%%%%%%%%%%%%%%%%%%%%%%%%%%%%%

\section{Contributions to the hypothesis of the
percolation breakdown\label{sec:hypothesis}}

In their study of RSA of needles on a square lattice,
Kondrat {\em et al.\@} \cite{Kondrat2001} noticed a peculiar dependence of the 
ratio of the percolation threshold to the jamming threshold 
($c_\mathrm{p}/c_\mathrm{j}$) on the needle length ($k$),
\begin{equation}
  \label{eq:log-kondrat}
     c_\mathrm{p}/c_\mathrm{j} \approx a + b \log_{10} k,
\end{equation}
with $a = 0.50$ and $b = 0.13$.
Since $c_\mathrm{p}/c_\mathrm{j}$ cannot be greater
than 1, it was clear that this relation
must break for needles of length larger than some characteristic length $k_*$,
which in this case can be estimated as
$10^{(1-a)/b} \approx 7000$. Equation~(\ref{eq:log-kondrat}) was put forward as a
phenomenological formula based on numerical results for needles
of rather moderate length $k \le 45$, which raised the question: is $k_*$ a
real physical parameter and if so, what happens to $c_\mathrm{p}/c_\mathrm{j}$ 
as $k$ approaches and then exceeds $k_*$?

This problem was tackled by Tarasevich {\em et al.\@} \cite{Tarasevich2012}, who
studied the RSA of partially ordered needles. In the isotropic case they confirmed
relation (\ref{eq:log-kondrat})
for much longer needles ($k \le 512$) and also obtained more
accurate estimates of parameters $a=0.513(6)$ and $b=0.119(3)$, which implies
$k_*=12\,400(3700)$.
As the logarithmic formula was verified for really long needles, the problem of
what happens close to and beyond $k_*$ became more interesting.
Assuming that (\ref{eq:log-kondrat}) is valid up to
$k_*$, they formulated the hypothesis that for $k\gtrsim k_*$
the system does not percolate and the ratio $c_\mathrm{p}/c_\mathrm{j}$ 
simply becomes undefined, which would solve the paradox.

This rather surprising conclusion was confirmed in a study of random sequential
adsorption of needles in imperfect systems \cite{Tarasevich2015}.
Two extensions of the original model were considered: either the needles have
some imperfect (nonconducting) segments (a so called $\mathrm{K}$ model) or the
lattice has some sites forbidden for adsorption ($\mathrm{L}$ model).
It turned
out that to each needle length $k$ corresponds a critical level of impurities
$d_*^\mathrm{K}$ or $d_*^\mathrm{L}$
(for models $\mathrm{K}$ and $\mathrm{L}$, respectively), above which no
percolation can be observed. They were found to verify the following
phenomenological relations
\begin{equation}
 \label{eq:model-L}
  d_*^\mathrm{L}(k)=a_\mathrm{L}
  \frac{(k_*^\mathrm{L})^{\alpha}-k^{\alpha}}{b+k^{\alpha}},
\end{equation}
where $a_\mathrm{L}$, $\alpha$, $b$, and
$k_*^\mathrm{L}$ are some fitting parameters (model L of imperfect
lattice), and
\begin{equation}
 \label{eq:model-K}
   d_*^\mathrm{K}(k)=a_\mathrm{K} \log_{10} (k_*^\mathrm{K}/k),
\end{equation}
with some  fitting parameters  $a_\mathrm{K}$ and $k_*^\mathrm{K}$
(imperfect needles).
For $k>k_*^\mathrm{L}$ (imperfect lattice) or $k > k_*^\mathrm{K}$ (imperfect
needles) these formulas predict an unphysical, negative value of the
critical impurity level. Moreover, the obtained values of $k_*^\mathrm{L}=5900(500)$
and $k_*^\mathrm{K}=4700(1000)$ are consistent with the previous estimates of $k_*$.
Therefore Tarasevich {\em et al.\@} concluded that their data
confirm the hypothesis of percolation breakdown
for sufficiently long needles.

Another method of introducing imperfections to the original model of random
adsorption of needles was recently
investigated by Centres {\em et al.\@} \cite{Centres2015}.
To model adsorption on surfaces of amorphous solids, they assumed
that all sites of the lattice are ready for adsorption, but a fraction $\rho$
of the bonds has been disabled before the adsorption begins.
They found a similar phenomenon to that found for lattice or needle impurities:
there exists a certain critical, $k$-dependent concentration of disabled bonds,
$\rho_*$, above which percolation does not occur. Moreover, their data fitted well
to (\ref{eq:model-L})
with the critical length $k_*=5518(500)$, a value consistent with previous reports.

%%%%%%%%%%%%%%%%%%%%%%%%%%%%%%%%%%%%%%%%%%%%%%%%%%%%%%%%%%%%%%%%%%%%%%%%%%%%%%
%%%%%%%%%%%%%%%%%%%%%%%%%%%%%%%%%%%%%%%%%%%%%%%%%%%%%%%%%%%%%%%%%%%%%%%%%%%%%%
%%%%%%%%%%%%%%%%    SECTION 3: Proof            %%%%%%%%%%%%%%%%%%%%%%%%%%%%%%
%%%%%%%%%%%%%%%%%%%%%%%%%%%%%%%%%%%%%%%%%%%%%%%%%%%%%%%%%%%%%%%%%%%%%%%%%%%%%%
%%%%%%%%%%%%%%%%%%%%%%%%%%%%%%%%%%%%%%%%%%%%%%%%%%%%%%%%%%%%%%%%%%%%%%%%%%%%%%

\section{Theorem and its proof\label{sec:proof}}

We will prove the following theorem:
\textit{Every jammed configuration of fixed-length nonoverlapping horizontal or
vertical needles on a finite square lattice contains a connected cluster spanning
two opposite edges of the lattice.}

In this theorem, a ``needle'' is a rectangle of size 1 lattice unit (l.u.) by $k$
l.u., with its corners located at the underlying lattice nodes. Two
needles are  connected directly if they share a part of their sides of length $\ge
1$ l.u.\@ so that for $k=1$ the problem reduces to the classical site
percolation. Of course, we consider only systems large enough to accommodate
at least one needle. Moreover, since the theorem is trivial for $k=1$,
henceforth we assume that $k>1$ and one can divide the needles into horizontal and
vertical ones.

Below we present two different methods of proving this theorem, as each of them can
potentially be used in more general cases e.g., for lattices other than the square
one \cite{Cornette2003b} or in higher space dimensions \cite{Tarasevich2007}.

%%%%%%%%%%%%%%%%%%%%%%%%%%%%%%%%%%%%%%%%%%%%%%%%%%%%%%%%%%%%%%%%%%%%%%%%%%%%%%
%%%%%%%%%%%%%%%%%%%%%%%   Method I     %%%%%%%%%%%%%%%%%%%%%%%%%%%%%%%%
%%%%%%%%%%%%%%%%%%%%%%%%%%%%%%%%%%%%%%%%%%%%%%%%%%%%%%%%%%%%%%%%%%%%%%%%%%%%%%

\subsection{Method I}
For convenience, we start from proving the following Lemma:
\textit{Every cluster at a jammed configuration extends
to one of two consecutive edges of the system.} If one
labels the system edges using the geographical notation (N, E, S, and W for
the top, right, bottom, and left edge, respectively), the lemma states
that \textit{any} cluster  at jamming must touch at least one edge in each of
the four pairs: (N, E), (E, S), (S, W) and (W, N).

We will prove the Lemma {\em a contrario}---let us assume that there exists
a jammed configuration of fixed-length nonoverlapping needles with a cluster
that does not touch any of two consecutive system edges, say N and E.

\begin{figure}
\begin{center}
\includegraphics[width={0.8\columnwidth}]{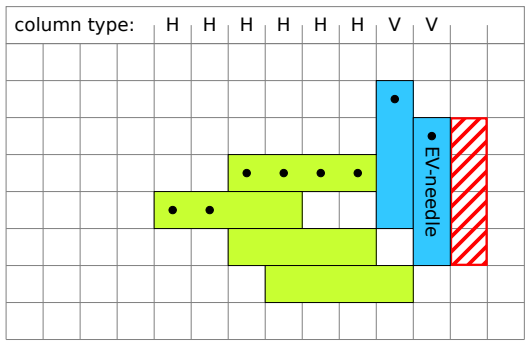}
\caption{An exemplary cluster with the topmost cell in each lattice column marked
with a dot. Orientation (horizontal or vertical) of the needles occupying these
cells can be used to classify the corresponding lattice columns as H or V.
If the rightmost column is of type V and the rightmost needle
(``EV-needle'') does not touch the edge of the system, there is a room, marked with
a hatched pattern, for another needle.}
\label{fig:x1}
\end{center}
\end{figure}

Since the system is a square lattice of size $L\times L$ ($L\ge k$), it can be
regarded as a set of $L$ columns, each made of $L$ elementary lattice cells.
For each column we can identify the set of all its cells that belong
to the cluster.
If this set is nonempty, we can use the topmost cell from this set to classify
the column as follows:
if the column's topmost cell belongs to a horizontal needle, the column is said
to be of type H,  otherwise the cell belongs to a vertical needle and the column
is said to be of type V. This is illustrated in Fig.~\ref{fig:x1}.

Consider now the rightmost column  containing the sites
from our cluster. We will label it $c_\mathrm{HE}$ because it is cluster's
easternmost column and it is of type H.
To see the reason for the latter property, suppose this column is of type
V. In such a case it would contain at least one vertical needle (marked as
an EV-needle in Fig.~\ref{fig:x1}).
Since we assumed that the cluster does not touch system's edge E,
there exists a column to the right of the EV-needle
and it contains at least $k$ consecutive empty sites, see Fig.~\ref{fig:x1}.
This, however, contradicts the assumption that the system is jammed.

There are now two possibilities: either all columns with the sites from the
cluster are of the same type H or at least one
of these columns is of type V. We will consider each of these cases separately.

\subsubsection{Case A: all columns are of type H}
\begin{figure}
\begin{center}
\includegraphics[width={0.8\columnwidth}]{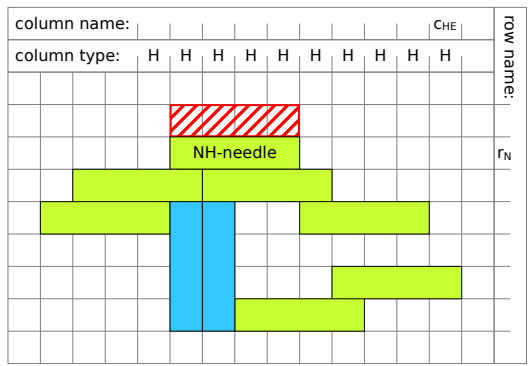}
  \caption{If all lattice columns are of type H
           and the cluster does not extend to the top edge of the system,
           there is a room (marked with a hatched pattern) for another needle
           directly above any horizontal needle in the cluster's topmost row,
$r_\mathrm{N}$.}
\label{fig:x2}
\end{center}
\end{figure}

Let $r_\mathrm{N}$ denote the topmost row containing the cluster (see
Figure~\ref{fig:x2}).
As all columns are assumed to be of type H, this row must contain at least one
horizontal needle, which we will call ``NH-needle''.
As no cell above the topmost row can belong to the cluster and since
we have assumed that the cluster does not touch system edge N, all cells neighboring
the NH-needle from above are empty and can accommodate another needle. As this
contradicts the assumption that the system is jammed, the proof of case A is
completed.

\subsubsection{Case B: columns of mixed types, H and V}
\begin{figure}
\begin{center}
\includegraphics[width={0.8\columnwidth}]{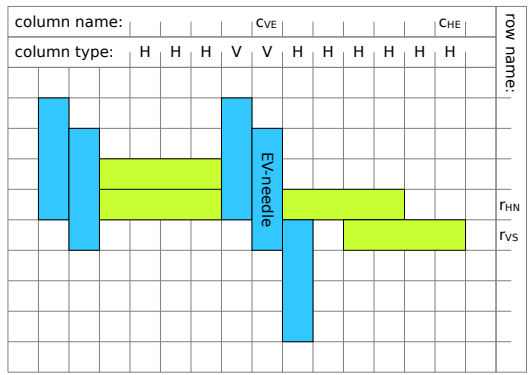}
\caption{For clusters occupying columns of type H and V,
         the ``EV-needle'' is defined as the topmost vertical needle in the
         rightmost column of type V,  $r_\mathrm{VS}$ is
         the lowest row occupied by this needle,
         and  $r_\mathrm{HN}$ is the topmost row
         containing a horizontal needle to the right of the EV-needle.
         }
\label{fig:x3}
\end{center}
\end{figure}

Consider now the case where at least one column is of type V.
Let $c_\mathrm{VE}$ denote the rightmost column of type V.
Cluster's topmost cell in this column belongs to a vertical needle,
which we call the EV-needle (see Fig.~\ref{fig:x3}).
Its bottom cell defines a reference row, which we denote as
$r_\mathrm{VS}$.
We also define $r_\mathrm{HN}$ as the topmost row containing a horizontal
needle in any of the columns located to the right of
column $c_\mathrm{VE}$. The remaining part of the proof depends on the
relation between $r_\mathrm{VS}$ and $r_\mathrm{HN}$.

\begin{figure}
 \begin{center}
  \includegraphics[width={0.8\columnwidth}]{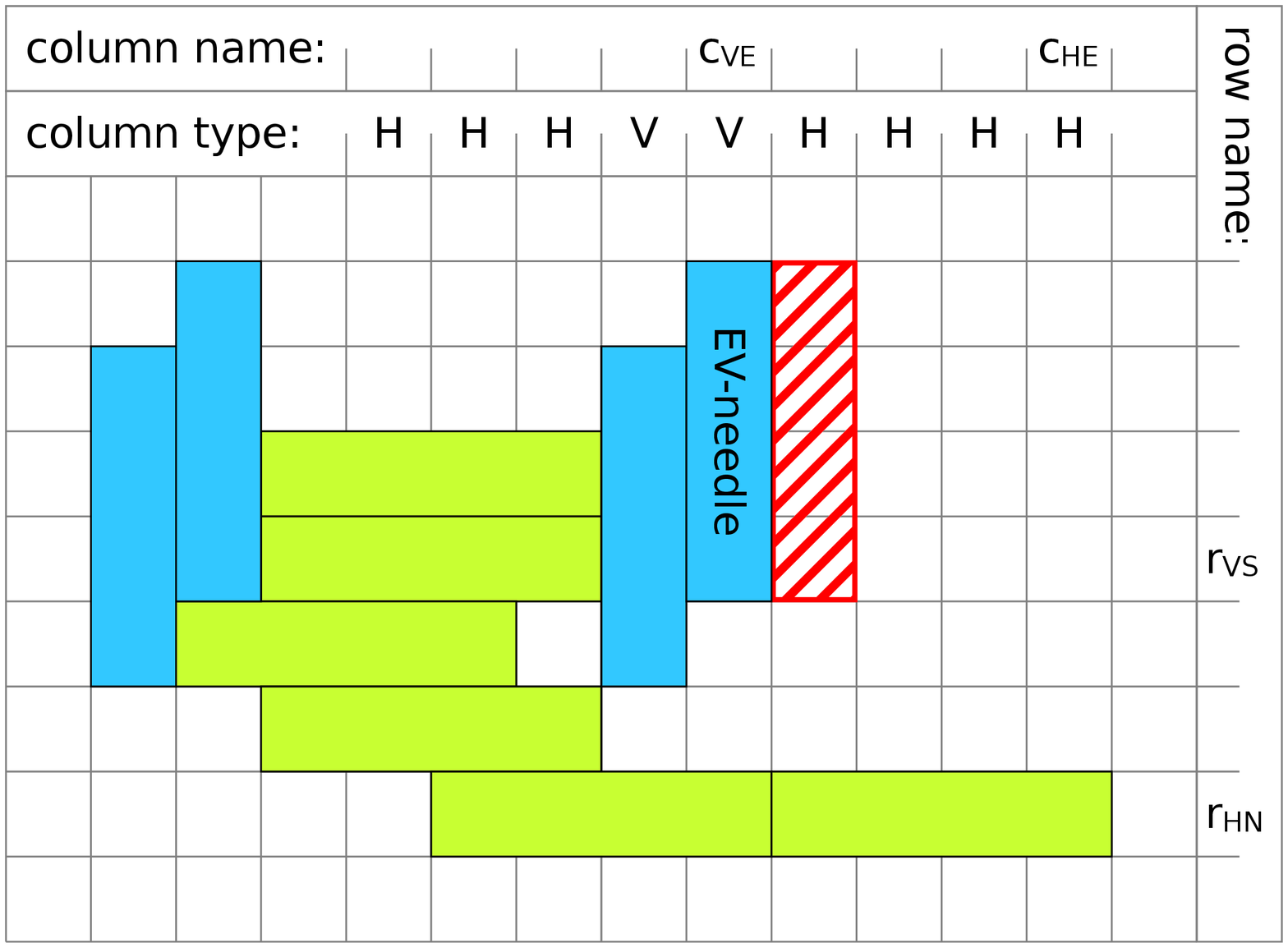}
  \caption{If $r_\mathrm{VS} > r_\mathrm{HN}$ and the cluster does not extend to the
           right edge of the system, the cells bordering the EV-needle
           from the right are empty and can hold another needle.}
  \label{fig:x4}
 \end{center}
\end{figure}

\paragraph{Case B1: $r_\mathrm{VS} > r_\mathrm{HN}$.}
In this case all cells of the cluster that are to the right of column
$c_\mathrm{VE}$ lie below row $r_\mathrm{VS}$.
This means that the cells located directly to the right of the EV-needle
are unoccupied and can accommodate another needle, see Fig.~\ref{fig:x4}.
This, however, contradicts the assumption that the system is jammed.

\begin{figure}
 \begin{center}
  \includegraphics[width={0.8\columnwidth}]{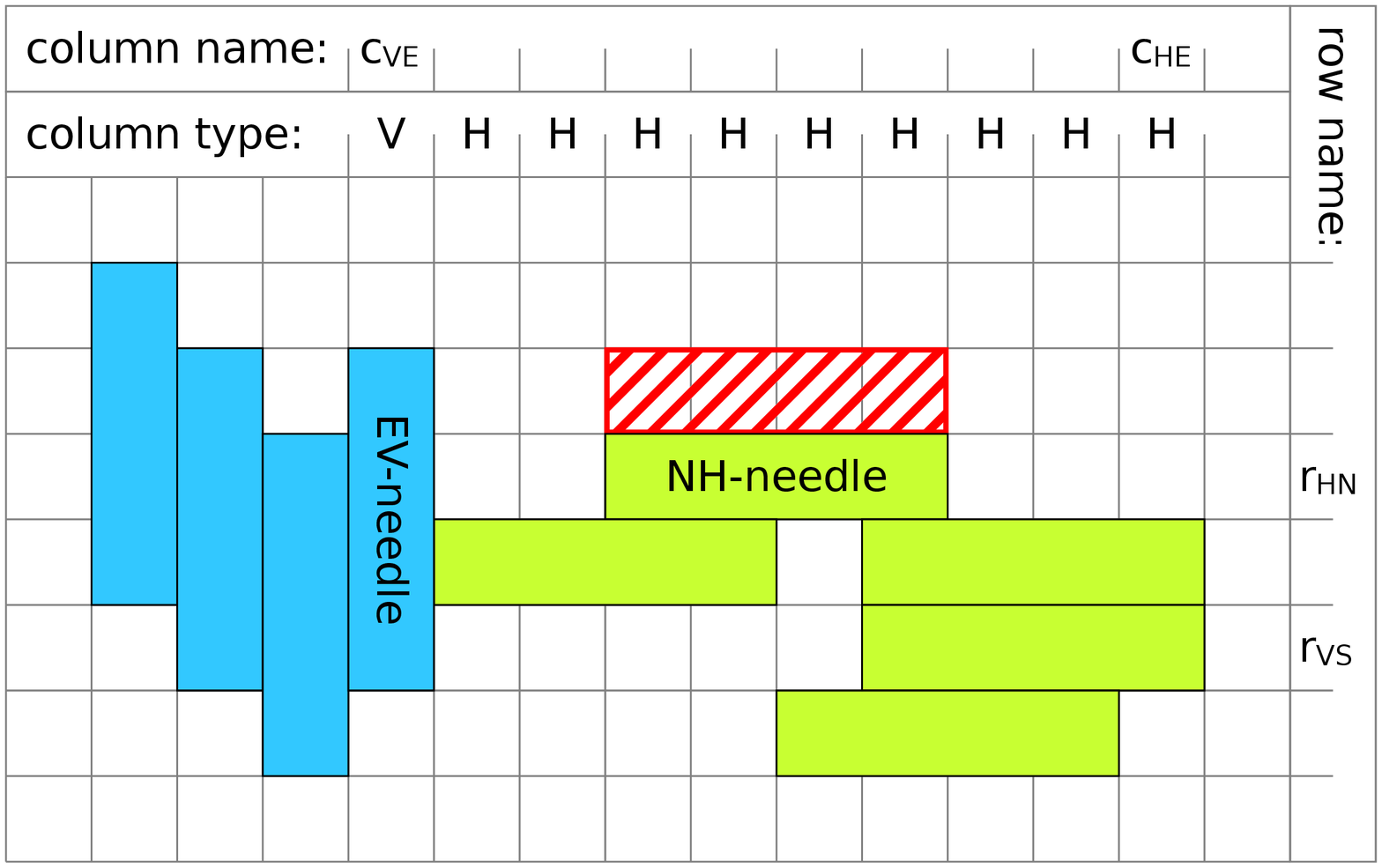}
  \caption{If $r_\mathrm{VS} \leq r_\mathrm{HN}$, there is a room for another
   needle directly above the NH-needle (the topmost horizontal
   needle to the right of the EV-needle).}
  \label{fig:x5}
 \end{center}
\end{figure}

\paragraph{Case B2: $r_\mathrm{VS} \leq r_\mathrm{HN}$.}
Let ``NH-needle'' denote the rightmost horizontal needle located at row
$r_\mathrm{HN}$.
Each cell occupied by this needle lies to the right of the EV-needle (see
Fig.~\ref{fig:x5}).
Directly above the NH-needle there are neither vertical needles (all columns
to the right of column $c_\mathrm{VE}$ are of type H), nor horizontal ones
(otherwise the NH-needle would not be the topmost one).
Thus, the cells neighboring the NH-needle from above are empty
and can accommodate another needle.
However, this contradicts the assumption that the system is jammed. This completes
the proof of the Lemma.

The final step is to show that the Lemma implies the Theorem. Let us assume that
the there is a jammed configuration of needles and a cluster in it. There are
two cases: either this cluster extends to all four edges of the
system or not. In the former case the cluster is trivially a percolating cluster.
In the latter one it does not touch at least one system edge, say, N. However, the
Lemma ensures that in this case it must extend to the two edges adjacent to
N, that is, to E and W, and hence must be a percolating cluster.

%%%%%%%%%%%%%%%%%%%%%%%%%%%%%%%%%%%%%%%%%%%%%%%%%%%%%%%%%%%%%%%%%%%%%%%%%%%%%%
%%%%%%%%%%%%%%%%%%%%%%%   Method II     %%%%%%%%%%%%%%%%%%%%%%%%%%%%%%%%
%%%%%%%%%%%%%%%%%%%%%%%%%%%%%%%%%%%%%%%%%%%%%%%%%%%%%%%%%%%%%%%%%%%%%%%%%%%%%%

\subsection{Method II}

Suppose that it is possible to fill a finite square lattice of size $L \times
L$, $L \ge k$, with nonoverlapping
horizontal or vertical needles of size
$1\times k$ in such a way that the system is jammed and a cluster
of connected needles exists such that it does not touch the system borders.
We will show that this assumption leads to a contradiction.

\begin{figure}
	\begin{center}
		\includegraphics[width={0.75\columnwidth}]{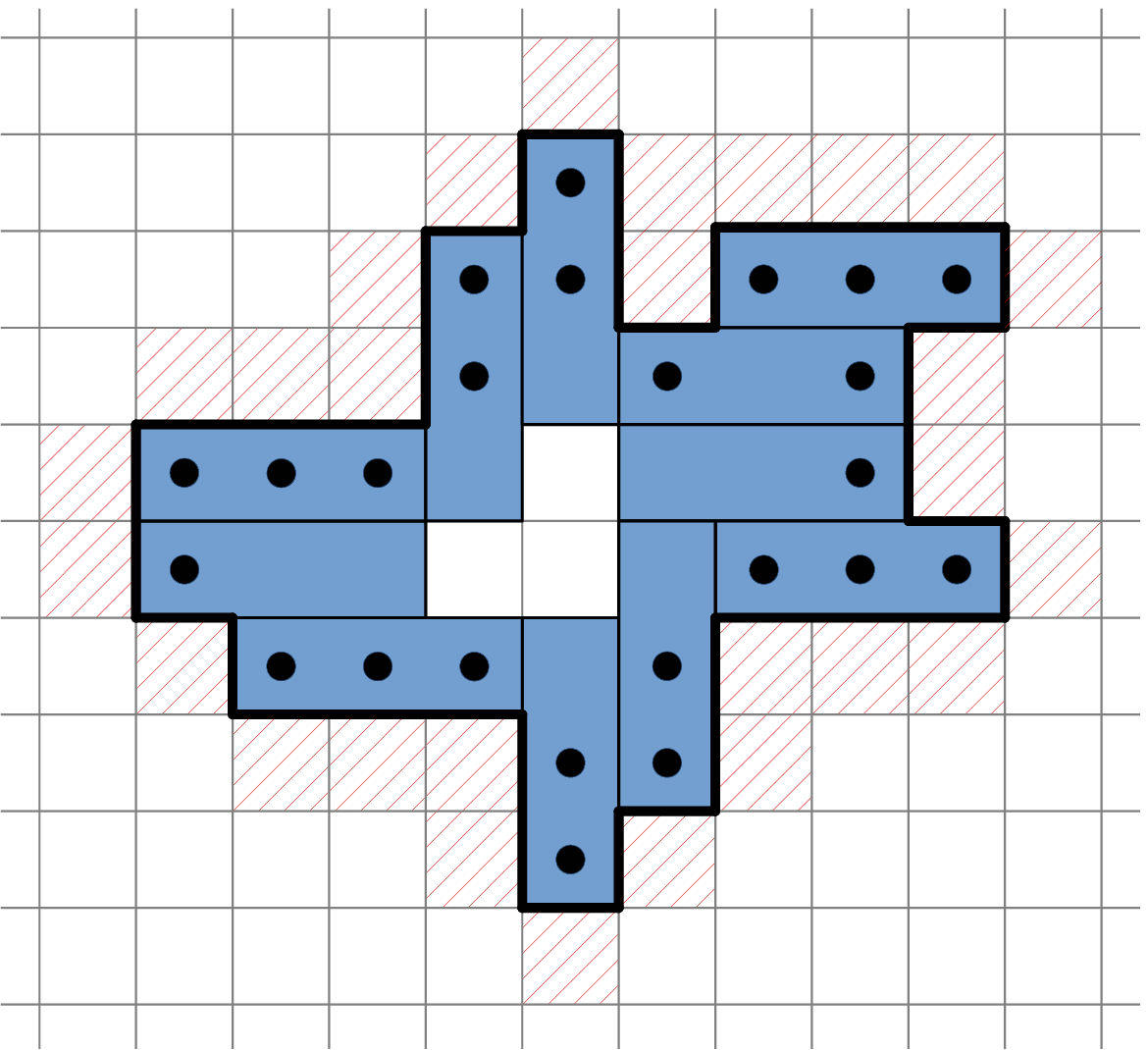}
		\caption{A hull (solid line) is a polygon composed of alternating
			horizontal and vertical sides tightly surrounding the needles forming a 
			cluster.
			The solid circles
			and the hatched pattern mark the lattice cells bordering the hull from 
			the inside
			and outside, respectively;
			by construction, the former are occupied by needles whereas the latter
			are empty.}
		\label{fig:6:zk}
	\end{center}
\end{figure}

Let us define the hull of a cluster as the minimal simple 
polygon encompassing it, see Fig.~\ref{fig:6:zk}.
This polygon is made of horizontal and vertical line segments of integer length.
It divides the plane into the space occupied by the cluster together
with, perhaps, some holes between the needles forming it,
and the remaining space.

By construction, any elementary square bordering the hull from inside must
be occupied by some needle, whereas none of the squares
bordering the hull from the outside can be occupied by a needle, see
Fig.~\ref{fig:6:zk}.
Thus, the length of any side of the hull is an integer less than $k$,
otherwise one could add
a needle on the squares bordering this side from the outside,
which contradicts the assumption that the system is jammed.

Each horizontal line segment of the hull is followed by a vertical one and so 
forth by turns, so that the number of its vertices, $N$, is even and the angles 
at its vertices are either $\pi/2$ or $3\pi/2$.
Let $q_+$ and $q_-$ denote the number of $\pi/2$ and $3\pi/2$ angles, 
respectively. Since the sum of internal angles of an $N$-sided polygon is 
$(N-2)\pi$, these quantities satisfy $q_+ + q_- = N$ and $q_+ \pi/2 + 
q_-(3\pi/2) = (N-2)\pi$. Consequently, 
\begin{equation}
 \label{eq:qq}
   q_+ = N/2+2, \quad q_- =N/2-2.
\end{equation}

\begin{figure}
\begin{center}
\includegraphics[width={0.95\columnwidth}]{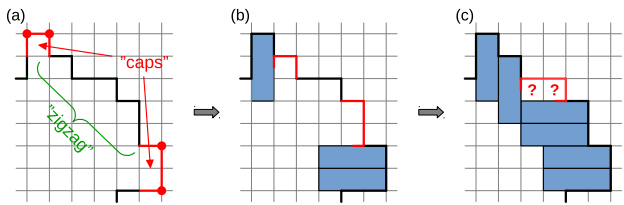}
\caption{The main ideas of Method II.
(a) In each hull one can find two ``caps'' (parts of the hull with two consecutive
right angles separated, perhaps, by a ``zigzag'' (part od a hull with consecutive
external and internal right angles); (b)~in a jammed state the length of each of
the hull's sides is smaller than that of a needle (here: $k=3$), which
uniquely determines the orientation of the needles filling in the caps;
(c)~the orientation of the needles touching the zigzag
must be the same as that in each cap, which leads to a contradiction, as they are
perpendicular to each other.
}
\label{fig:7:zk}
\end{center}
\end{figure}

Let us assign to each vertex $i$ of the hull an integer $a_i\in\{-1,1\}$ such 
that $a_i = 1$ if the internal angle at $i$ is $\pi/2$ and $a_i=-1$
if this angle is $3\pi/2$. The idea is that as we walk along the hull in a 
clockwise direction, we keep track of our current orientation by adding $1$ 
whenever we make a right turn and subtracting $1$ for the left turn. 
From (\ref{eq:qq}) we have
\begin{equation}
   \sum_{i=0}^{N-1} a_i = q_+ - q_- =  4,
\end{equation}
that is, whenever we return to the starting point, we must have made 4 more 
right turns than the left ones. 
Let $s_i= (a_i+a_{i+1})/{2}$ (throughout the paper we apply to the indices the 
modular arithmetic with modulus $N$).
By construction, $s_i$ are also integers, $s_i\in\{-1,0,1\}$, which measure the 
effect of making two consecutive turns. They satisfy 
\begin{equation}
   \sum_{i=0}^{N-1} s_i = \sum_{i=0}^{N-1} a_i = 4.
\end{equation}
In the sequence $s_0, s_1,\ldots,s_{N-1}$ there are
thus at least two indices $j < k$ such that $s_j = s_k = 1$ and
$s_l = 0$ for all $j < l < k$.
Therefore there exist two vertices, $j$ and $k$,
such that  $a_j = a_{j+1}= 1$ and $a_k = a_{k+1} = 1$,
separated, perhaps,  by an alternating sequence of the form $-1,1,\ldots,-1$.
Geometrically this corresponds to a ``zigzag'' ended by two ``caps'',
see Fig.~\ref{fig:7:zk}a.

Each cap uniquely determines the orientation
(horizontal or vertical) of the needles that occupy it.
This is because the length of the cap  side connecting two consecutive right angles
is smaller than $k$, and so any needle touching it from inside
must be perpendicular to it (cf.\@~Fig.~\ref{fig:7:zk}b).
Moreover, if the zig-zag segment exists between the two caps, each of the caps
enforces the direction of the needles touching the zig-zag segment
to be parallel to the needles filling in that cap.
However, the orientations of the two consecutive caps are orthogonal to each other.
This leads to a contradiction: the needles touching the zig-zag
cannot be all both horizontal and vertical (Fig.~\ref{fig:7:zk}c).

If, however, no zig-zag part exists,
then $k$ = $j+1$ and $a_j=a_{j+1} = a_{j+2} =1$,
that is, the cap has two consecutive, orthogonal sides of size smaller than 
$k$. Such a region cannot be filled by needles of length $k$,
which again contradicts our assumptions.
Thus, either the hull of the cluster touches one of the edges of a finite system,
in which case at least one side touching the edge
is of length $\ge k$, or the system is not jammed.

\begin{figure}
  \begin{center}
\includegraphics[width={0.77\columnwidth}]{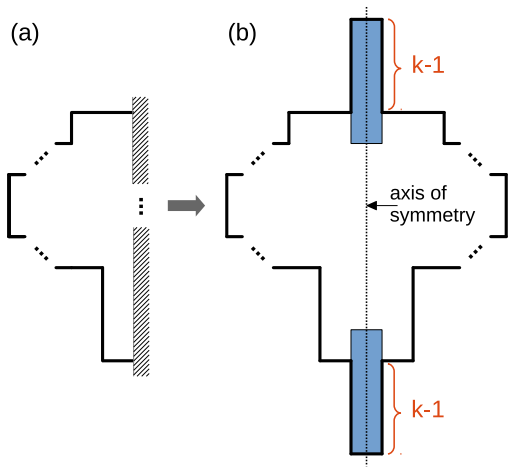}
\caption{(a) The hull of a hypothetical cluster touching only one edge of the
system. The lengths of all of its sides not touching the edge must be smaller 
than the needle length, $k$. 
(b) Construction of another cluster with the hull such 
that the lengths of all its sides would be smaller than $k$.
}
\label{fig:8:zk}
\end{center}
\end{figure}

Is it possible that a cluster at a jamming state touches
only one of the system's edges?
If such a cluster existed, it could be used to construct a cluster made
of needles of size $1\times k$, whose hull is a polygon with all sides of 
lengths~$<k$, see Fig.~\ref{fig:8:zk}.
However, we have just proven that such a cluster does not exist.
The construction is defined as follows. One takes the original cluster 
as well as its mirror reflection and joins them together with two needles 
sticking out with $k-1$ of their $k$ elementary segments. 
Each side of the resulting cluster would be smaller than $k$. 
Moreover, the two extra needles cannot overlap, for  
the original cluster must touch the system edge at at least
$k\ge2$ different lattice cells (corresponding to a longer side of one of its
needles). If this construction generates one or more holes inside the new
hull, one can fill them up with additional needles in an arbitrary way.

Thus any cluster at a jamming state must touch at least two edges of the system. If
a cluster touches exactly two consecutive edges of the system, one could use a
construction similar to that described in the previous paragraph twice, once in
the horizontal and once in the vertical direction,
to construct a cluster made of needles $1\times k$, whose
hull is a polygon with all sides of lengths $<k$. However, such a cluster does not
exist. Consequently, any cluster at a jamming state connects two opposite sides of
the system, horizontally or vertically, and therefore is a percolating cluster.

%%%%%%%%%%%%%%%%%%%%%%%%%%%%%%%%%%%%%%%%%%%%%%%%%%%%%%%%%%%%%%%%%%%%%%%%%%%%%
%%%%%%%%%%%%%%%%%%%%%%%%%%%%%%%%%%%%%%%%%%%%%%%%%%%%%%%%%%%%%%%%%%%%%%%%%%%%%
%%%%%%%%%%%%%%%%%%%%%%%   SECTION: CONCLUSIONS    %%%%%%%%%%%%%%%%%%%%%%%%%%%
%%%%%%%%%%%%%%%%%%%%%%%%%%%%%%%%%%%%%%%%%%%%%%%%%%%%%%%%%%%%%%%%%%%%%%%%%%%%%
%%%%%%%%%%%%%%%%%%%%%%%%%%%%%%%%%%%%%%%%%%%%%%%%%%%%%%%%%%%%%%%%%%%%%%%%%%%%%

\section{Conclusions and outlook\label{Sec:Conclusions}}

We have proved that all jammed configurations of 
nonoverlapping needles of size $1\times k$ ($k$-mers) on a 
square lattice are percolating ones. 
This disproves  the recent conjecture 
\cite{Tarasevich2012,Tarasevich2015,Centres2015} that in the random sequential 
adsorption of such needles on a square lattice the 
percolation does not occur if the needles are longer than some threshold value  
$k_*$, estimated to be of order of several thousand.

While this result ensures that the percolation to jamming ratio  
($c_\mathrm{p}/c_\mathrm{j}$) is well defined for all needle lengths, 
it does not bring us much closer to the understanding of how this ratio
varies with $k$ for $k\gtrsim 500$. 
Perhaps the only way of obtaining this information is through numerical 
simulations, but this would require either to employ supercomputers or to 
devise much more efficient algorithms tailored to this specific problem. 

Our theorem has some implications for other RSA problems. For 
example, in the case of the RSA on an imperfect lattice \cite{Tarasevich2015}, 
we can use it to conclude that any cluster formed in the jammed, 
nonpercolating state must have at least two ``nonconducting'' lattice cells at 
its perimeter, both adjacent to the sides of the hull whose length is $\ge k$. 
This helps to understand why even a tiny lattice impurity can preclude 
percolation in systems with very long needles: one impurity per cluster hull 
side can be enough to stop its growth.

Another interesting point is whether our 
arguments can be extended to other lattices, e.g. the triangular or cubic ones.

%%%%%%%%%%%%%%%%%%%%%%%%%%%%%%%%%%%%%%%%%%%%%%%%%%%%%%%%%%%%%%%%%%%%%%%%%%%%%%
%%%%%%%%%%%%%%%%%%%%%%%%%%%%%%%%%%%%%%%%%%%%%%%%%%%%%%%%%%%%%%%%%%%%%%%%%%%%%%
%%%%%%%%%%%%%%%%%%%%%%%%%%%    REFERENCES    %%%%%%%%%%%%%%%%%%%%%%%%%%%%%%%%%
%%%%%%%%%%%%%%%%%%%%%%%%%%%%%%%%%%%%%%%%%%%%%%%%%%%%%%%%%%%%%%%%%%%%%%%%%%%%%%
%%%%%%%%%%%%%%%%%%%%%%%%%%%%%%%%%%%%%%%%%%%%%%%%%%%%%%%%%%%%%%%%%%%%%%%%%%%%%%

\section*{References}

\end{document}